\documentclass[submission,copyright,creativecommons]{eptcs}
\providecommand{\event}{FROM 2024} 

\usepackage{iftex}
\usepackage{textcomp}

\ifpdf
  \usepackage{underscore}         
  \usepackage[T1]{fontenc}        
\else
  \usepackage{breakurl}           
\fi

\usepackage{listings}
\usepackage{cite}
\usepackage{algorithmic}
\usepackage{textcomp}
\usepackage{xcolor}
\def\BibTeX{{\rm B\kern-.05em{\sc i\kern-.025em b}\kern-.08em
    T\kern-.1667em\lower.7ex\hbox{E}\kern-.125emX}}
    
\usepackage{graphicx}
\usepackage[utf8]{inputenc}
\usepackage{amsmath}
\usepackage{amssymb}
\usepackage{stmaryrd}
\usepackage{tikz}
\usepackage{multirow}
\usetikzlibrary{arrows,positioning,shapes,calc,tikzmark,patterns,snakes}
\usepackage{comment}
\usepackage{versions}
\usepackage{algorithm2e}
\usepackage{ upgreek }
\usepackage{wasysym}
\usepackage[normalem]{ulem}
\usepackage{hyperref}
\usepackage{float}
\usepackage{adjustbox}

\usepackage[font=small,labelfont=bf,tableposition=top]{caption}
\DeclareCaptionLabelFormat{andtable}{#1~#2  \&  \tablename~\thetable}
\usepackage{enumitem}

\usepackage{algorithmic}

\includeversion{ICALP} 
\excludeversion{ARXIV} 

\newcommand{\DNK}{\textnormal{DyNetKAT}}

\newcommand{\DNA}{\DNK}

\newcommand{\NetKATnoDup}{\textnormal{NetKAT}^{-{\bf{dup}}}}

\newcommand{\drop}{\mathbf{0}}

\newcommand{\cpolX}{cpol_{X}}

\newcommand{\cpolseqsucc}{cpol_{\_\Seq}^{\checkmark}}

\newcommand{\cpoloplusl}{cpol_{\_\oplus}}

\newcommand{\Symb}[1]{\bf{Symb}_{#1}}

\newcommand{\rec}[1]{\mathbf{rcfg(#1)}}

\newcommand{\cpolmsg}{cpol_{\bullet}}

\newcommand{\reconfig}{cpol_{\clubsuit \spadesuit}}

\newcommand{\intl}{cpol_{\_||}}

\newcommand{\packcopy}{\mathbf{1}}
\newcommand{\zeroq}{\bot}

\newcommand{\Par}{\mathop{||}}
\newcommand{\Seq}{\mathop{;}}

\newcommand{\xgtrans}[2]{\,\,{{\xrightarrow{#2}}_{#1}}\,\,}

\newcommand{\trans}[1]{\xgtrans{}{#1}}

\newcommand{\sosrule}[2]{\frac{\raisebox{.7ex}{\satsize{$#1$}}}
                        {\raisebox{-1.0ex}{\satsize{$#2$}}}}
\newcommand{\satsize}{\normalsize}

\newcommand{\dedr}[1]{\ensuremath{\mathbf{(#1)}}}

\newenvironment{todo}{\bigskip\hrule\medskip\noindent}{\medskip\hrule\bigskip}

\newcommand{\pushright}[1]{\ifmeasuring@#1\else\omit\hfill$\displaystyle#1$\fi\ignorespaces}
\newcommand{\pushleft}[1]{\ifmeasuring@#1\else\omit$\displaystyle#1$\hfill\fi\ignorespaces}
\usepackage{enumitem}

\definecolor{mygreen}{rgb}{0.0, 0.5, 0.0}

\usepackage{graphicx}

\title{Tracer: A Tool for Race Detection in\\ Software Defined Network Models
}
\author{Georgiana Caltais
\institute{University of Twente\\ The Netherlands}
\email{g.g.c.caltais@utwente.nl}
\and
Mahboobeh Zangiabady
\institute{University of Twente\\ The Netherlands}
\email{\quad m.zangiabady@utwente.nl}
\and
Ervin Zvirbulis
\institute{University of Twente\\ The Netherlands}
\email{\quad e.zvirbulis@student.utwente.nl}
}
\def\titlerunning{Tracer}
\def\authorrunning{Caltais et al.}
\begin{document}
\maketitle

\begin{abstract}
Software Defined Networking (SDN) has become a new paradigm in computer networking, introducing a decoupled architecture that separates the network into the data plane and the control plane. The control plane acts as the centralized brain, managing configuration updates and network management tasks, while the data plane handles traffic based on the configurations provided by the control plane. Given its asynchronous distributed nature, SDN can experience data races due to message passing between the control and data planes. This paper presents Tracer, a tool designed to automatically detect and explain the occurrence of data races in DyNetKAT SDN models. DyNetKAT is a formal framework for modeling and analyzing SDN behaviors, with robust operational semantics and a complete axiomatization implemented in Maude. Built on NetKAT, a language leveraging Kleene Algebra with Tests to express data plane forwarding behavior, DyNetKAT extends these capabilities by adding primitives for communication between the control and data planes. Tracer exploits the DyNetKAT axiomatization and enables race detection in SDNs based on Lamport vector clocks. Tracer is a publicly available tool.
\end{abstract}

\section{Introduction}\label{sec:intro}

\begin{figure}[tb]
    \centering    \includegraphics[width=0.7\linewidth]{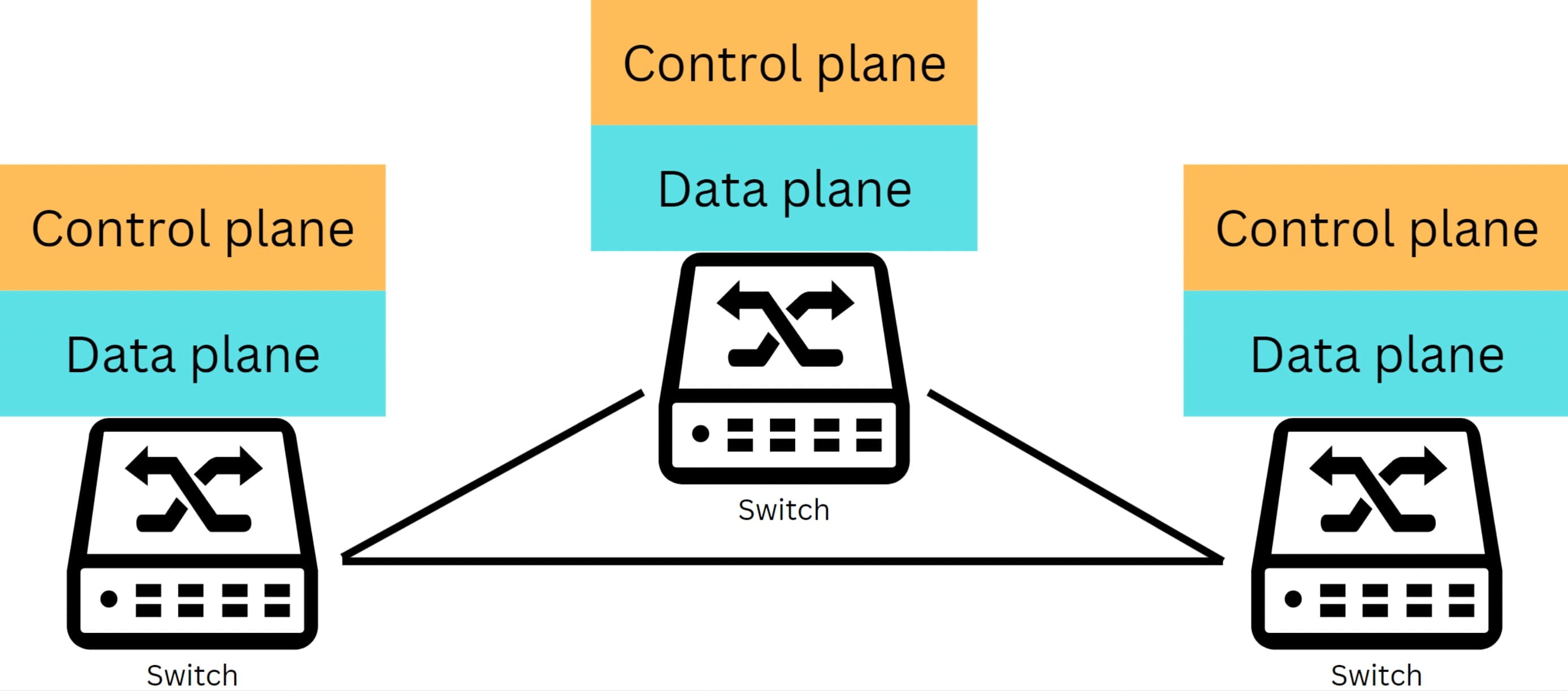}
    \caption{Traditional network setup}
    \label{fig:traditionalnetwork}
\end{figure}

Traditional network devices have been called “the last bastion of mainframe computing” \cite{hamilton2009networking}. Since the 1970s, network design principles have remained fundamentally unchanged, maintaining their core structure for nearly four decades. One of such fundamentals is the handling of the data and control planes. Intuitively, the data plane is a distinct functional layer in networking responsible for the forwarding of data packets between network devices. The control plane is another layer responsible for network control including policy enforcing and routing configuration. 
In a traditional network, each switch autonomously manages its interpretation of the control plane as illustrated in Figure~\ref{fig:traditionalnetwork}.
This architectural rigidity increases complexity in network maintainability due to the necessity of configuring each switch individually. 

\begin{figure}[h]
    \centering
    \includegraphics[width=0.7\linewidth]{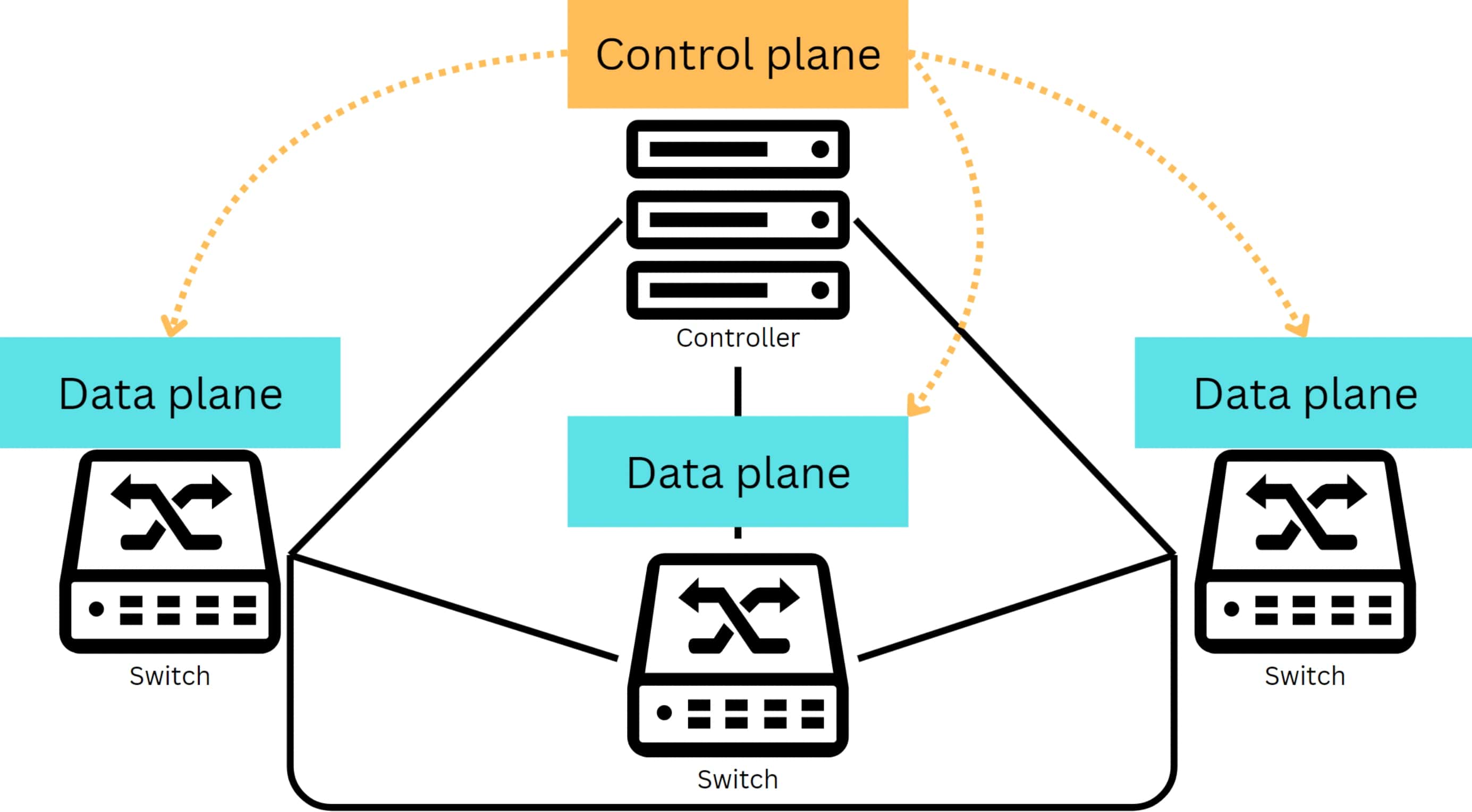}
    \caption{Software-defined network setup}
    \label{fig:SDN}
\end{figure}

In response, the concept of software-defined networking (SDN) has emerged.
The main difference is the separation of the data and control planes and consolidation of the management over the control plane in a centralized location as illustrated in Figure~\ref{fig:SDN}. 
SDN architectures comprise central controllers and programmable switches that communicate via standardized protocols. The former respond to network events such as new connections from hosts, topology changes, and shifts in traffic load, by re-programming the switches accordingly (as indicated by the orange dotted arrows in the figure). Such an approach enhances network controllability and adaptability in real-time scenarios. 
SDN is being adopted across various leading tech companies and cloud service providers to enhance the agility, efficiency, and scalability of their data center networks.
For instance, Google's B4 that connects Google's data centers across the world uses a centralised SDN controller that manages the entire network. Microsoft has also implemented SDN extensively in its Azure cloud infrastructure. 
Amazon employs SDN to support its scalable and flexible cloud services.

SDNs are highly concurrent systems, designed to handle numerous simultaneous operations. Consequently, they are prone to data races. The latter can lead to undesired outcomes/behaviours of the SDN, especially if the races correspond to concurrency between the data and the control planes.

\begin{figure}[tb]
    \centering    \includegraphics[width=0.7\linewidth]{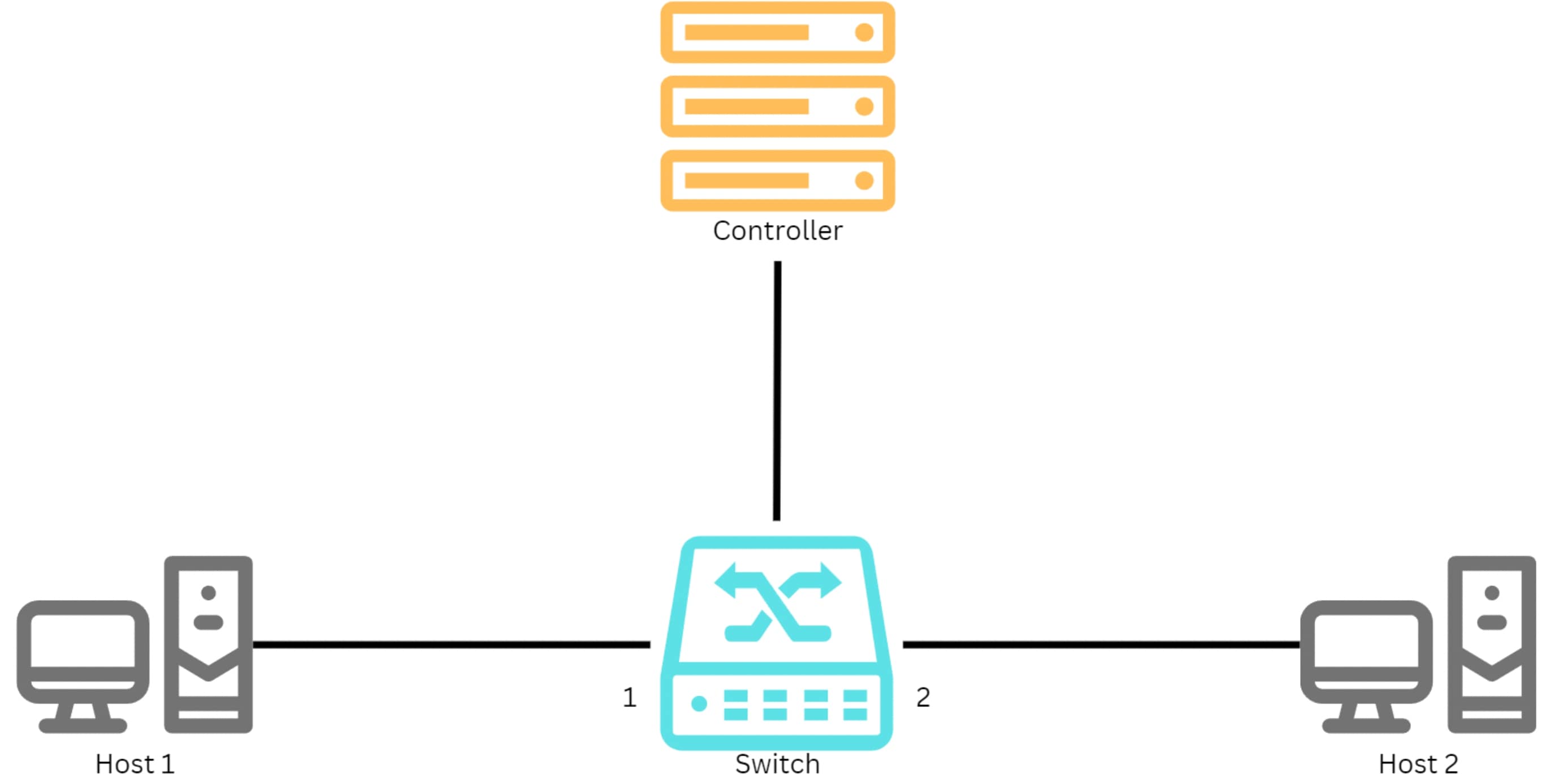}
    \caption{Running example inspired from~\cite{race-openflow}}
    \label{fig:SDN-race}
\end{figure}

Let us consider Figure~\ref{fig:SDN-race} (inspired from~\cite{SDN-CHF}), illustrating a basic example of an SDN consisting of: (i) a switch with two ports (1 and 2), (ii) one controller communicating with the switch, and (iii) two hosts (Host 1 and Host 2) that can send/receive packets to/from the switch via the aforementioned ports.
Assume the following over-simplified scenario: The switch is configured to allow any traffic from port 1 to port 2. When the switch encounters a ``blocking'' flag, it notifies the controller and continues forwarding subsequent packets until a new forwarding policy (``drop everything'', in this case) is received from the controller. If Host 1 sends a packet flagged ``blocking'' to the switch, a data race may occur. The race arises because the outcome for a new packet depends on the timing. The new packet will either be forwarded according to the existing forwarding policy installed in the switch, if it arrives before the blocking rule from the controller, or it will be dropped if the blocking rule is in place first.

In this paper, we propose Tracer~\cite{tracer}, a tool for the automated detection of data races in SDNs. Tracer builds around DyNetKAT~\cite{CaltaisHMT22}, a formal framework for the rigorous modelling and analysis of SDNs. DyNetKAT can encode and simulate packet forwarding within SDNs, together with the actual communication between the control and data planes (or, dynamic network reconfigurations).
The DyNetKAT language is supported by a rigorous operational semantics and a sound and complete axiomatisation enabling reasoning about equivalence of DyNetKAT programs, and associated packet forwarding behaviour in a fully automated fashion. Intuitively, Tracer takes as input DyNetKAT models and checks whether data races between the control and data planes occur, by following the symbolic approach in~\cite{SDN-CHF}. Furthermore, Tracer provides explanations of how such races can be enabled via (minimal) sequences of packets fed to the network. Such explanations can serve as a great debugging aid for the network administrators.

As mentioned in~\cite{SDN-CHF}, several methods have been developed to detect race conditions in SDNs, including ConGuard~\cite{XuHHZG17} and SDNRacer~\cite{El-HassanyMBVV16}. These tools identify race condition vulnerabilities by analyzing dynamically generated log files to construct an event graph where happens-before edges connect events, and race conditions manifest as partially ordered events. The Spin model checker, as discussed in~\cite{VinarskiiLKYZ19}, has also been employed to detect race conditions through runtime monitoring of events in SDNs.
In contrast, this paper is based on a \emph{static approach to identify races} in SDNs, eliminating the need for dynamic log generation from a network.

\medskip
\textit{Our contributions:} As previously mentioned, in this paper we introduce Tracer~\cite{tracer}, a tool for the automated detection of races in SDN models encoded in DyNetKAT, based on the theoretical framework in~\cite{SDN-CHF}. In short, Tracer exploits the symbolic semantics of DyNetKAT in~\cite{SDN-CHF} and uses Lamport vector clocks for detecting races entailed by the concurrent message passing between the SDN control and data planes as in~\cite{SDN-CHF}. Furthermore, Tracer provides explanations of how such races are enabled by computing minimal sets of network packets that lead to not well-behaved communication scenarios. The instructions for installing and running Tracer are publicly available in~\cite{tracer}.

\medskip
\textit{Structure of paper:} In Section~\ref{sec:dynetkat}, we briefly recall (Dy)NetKAT and introduce our running example. In Section~\ref{sec:vec-clocks}, we present the idea behind vector clocks for race detection in distributed systems. The symbolic semantics of DyNetKAT enriched with vector clocks is recalled in Section~\ref{sec:symb-dynetkat}. Our tool, Tracer, is introduced in Section~\ref{sec:tracer}. We draw the conclusions and provide pointers to future work in Section~\ref{sec:conc}.

\section{Overview of DyNetKAT}\label{sec:dynetkat}

DyNetKAT~\cite{CaltaisHMT22} serves as a framework for representing and analyzing the behaviors of SDNs, such as packet forwarding and the interaction between the control and data planes. DyNetKAT is an extension of NetKAT~\cite{netkat}, which is a language based on Kleene Algebra with Tests~\cite{DBLP:journals/toplas/Kozen97}, tailored for modeling and analyzing data plane forwarding. DyNetKAT introduces concurrency to NetKAT, in order to support dynamic reconfigurations of the data plane, such as the installation of new forwarding rules, in line with control plane protocols. 

Figure~\ref{fig:NetKAT-syn-sem} illustrates the syntax and semantics of the NetKAT language.
Network packets are encoded in (Dy)NetKAT as collections of fields, and associated values ranging over finite domains: $\{f_1 = v_1, \ldots, f_n = v_n\}$. For instance, a packet $\sigma$ of type $SSH$ residing at port $1$ of switch $SW_A$, with destination $Host_1$, can be conveniently denoted as $\sigma \triangleq \{type = SSH, pt = 1, SW = SW_A, dst = Host_1\}$.
The main syntactic elements of NetKAT include primitives $(Pr)$ for dropping incoming packets ($\drop$) and accepting incoming packets without further processing ($\packcopy$).
NetKAT primitives can filter out packets based on tests ($f = n$) and their disjunction ($+$), conjunction ($\cdot$) and negation ($\neg$). NetKAT policies $(N)$ can also be used for packet fields modifications ($f \leftarrow n$), or to express packet multicasting ($+$), composition of policies ($\cdot$) and iteration ($^*$). The operator $\textbf{dup}$ is designed for building histories of packets processed by an SDN dataplane encoded in NetKAT. The denotational semantics of NetKAT is defined over sets of packet histories as in Figure~\ref{fig:NetKAT-syn-sem}; an intuitive description of its operators has been provided earlier.
Furthermore, NetKAT has a sound and complete axiomatization that has been effectively used to reason about packet reachability within NetKAT models.

\begin{figure}[ht]
\[
\footnotesize{
\begin{array}{rcl}
 \multicolumn{3}{l}{\textnormal{\bf{NetKAT Syntax:}}} \\
 \mathit{Pr} & ::= & \drop \mid \packcopy{} \mid f = n \mid \mathit{Pr} + \mathit{Pr} \mid \mathit{Pr} \cdot \mathit{Pr} \mid \neg \mathit{Pr} 
 \\
     N & ::= & \mathit{Pr} \mid f \leftarrow n \mid N + N \mid N \cdot N \mid N^* \mid \textbf{dup}
\end{array}
}
\]
\[
\footnotesize{
\begin{array}{cc}
    \begin{array}{rcl}
   \multicolumn{3}{l}{\textnormal{\bf{NetKAT Semantics:}}} \\
   \llbracket \packcopy \rrbracket (h) & \triangleq & \{h\}\\
   \llbracket \drop \rrbracket (h) & \triangleq & \{\}\\
   \llbracket f=n \rrbracket \; (\sigma{:}{:}h) & \triangleq &\; \left\{
                \begin{array}{ll}
                  \{\sigma{:}{:}h\} & \textnormal{if}\; \sigma(f) = n \\
                  \{ \} & \textnormal{otherwise}
                \end{array}
              \right.\\
    \llbracket \neg a \rrbracket \; (h) & \triangleq &\; \{h\} \setminus\,\, \llbracket a \rrbracket \; (h) \\
    \llbracket f \leftarrow n \rrbracket \; (\sigma{:}{:}h) &\triangleq &\; \{ \sigma[f := n]{:}{:}h\} \notag \\
    \llbracket p + q \rrbracket \; (h) &\triangleq &\; \llbracket p \rrbracket \; (h) \cup \llbracket q \rrbracket \; (h)\\

 \end{array}
 &
  \begin{array}{rcl}
    \llbracket p \cdot q \rrbracket \; (h)  &\triangleq &\; ( \llbracket p \rrbracket \bullet \llbracket q \rrbracket) \; (h) \\
    \llbracket p^* \rrbracket \; (h) &\triangleq &\; \bigcup_{i \in N} F^i\;(h)  \\
F^0\; (h) &\triangleq &\; \{ h \}\\
F^{i+1}\;(h) &\triangleq &\; (\llbracket p \rrbracket \bullet F^i)\;(h)  \\
(f \bullet g)(x) &\triangleq &\; \bigcup \{g(y) \mid y \in f(x)\}\\
\llbracket \textbf{dup} \rrbracket \; (\sigma{:}{:}h) &\triangleq &\; \{\sigma{:}{:}(\sigma{:}{:}h)\} 

 \end{array}
\end{array}
}
\]
\caption{NetKAT: Syntax and Semantics~\cite{netkat}}
\label{fig:NetKAT-syn-sem}
\end{figure}

The syntax of {\DNA} is defined on top of the \textbf{dup}-free fragment of NetKAT as in~(\ref{def::DNA-syntax}). The constant $\bot$ denotes a DyNetKAT process without behaviour. Sequential composition of DyNetKAT policies $D$ is denoted by $\Seq$ . The operator $\Par$ encodes concurrent behaviours of DyNetKAT policies, whereas $\oplus$ stands for non-deterministic choice. (A)synchronous communication in DyNetKAT is modeled in an ACP~\cite{DBLP:books/daglib/0069083}-style via message sending operators $x!N\Seq D$ and receiving operators $x?N\Seq D$. Intuitively, messages $N$ (e.g., NetKAT flow tables) can be exchanged via channels $x$ as a result of the communication between the control and data planes. As soon as such a new forwarding policy $N$ is received via $x$, the continuation $D$ can update its behaviour according to $N$. This would correspond to installing a new forwarding policy $N$ in the dataplane. Variables $X$ enable defining recursive DyNetKAT policies.

\begin{equation}\label{def::DNA-syntax}
\small
\begin{array}{ccl}
     \mathit{N} & ::= & {\NetKATnoDup}\\[1ex]
     \mathit{D} & ::= &
     { \zeroq } \mid
     \mathit{N} \Seq \mathit{D}
     \mid  x?\mathit{N} \Seq D \mid 
     x!\mathit{N} \Seq D \mid \mathit{D} \Par \mathit{D} \mid
     D \oplus D \mid X  \\
     && X \triangleq D
     
\end{array}
\end{equation}

\begin{figure*}[t]
\centering
\begin{adjustbox}{max width=\dimexpr0.8\textwidth}
{
$
\begin{array}{|c|c|}
\hline &
\\
\dedr{\cpolseqsucc} \sosrule{{\sigma' \in \llbracket p \rrbracket(\sigma \!\!::\!\! \langle \rangle)} }{ ( p ; q, \sigma :: H, H')  \trans{(\sigma, \sigma')} ( q, H , \sigma' :: H')}
& 
\dedr{\cpolX}\sosrule{(p, {H_0,  H_1}) \trans{\gamma} (p', {H'_0,  H'_1}) }{ ( X, H_0, H_1)  \trans{\gamma} ( p' , H'_0, H'_1)} X \triangleq p
\\ & \\
\hline  &
\\
\dedr{\cpoloplusl}\sosrule{(p, H_0, H'_0)  \trans{\gamma} (p', H_1, H'_1)  }{ (p \oplus q, H_0, H'_0)  \trans{\gamma} ( p', H_1, H'_1)}
& 
\dedr{\intl}\sosrule{(p, H_0, H'_0)  \trans{\gamma} (p', H_1, H'_1)  }{ (p || q, H_0, H'_0)  \trans{\gamma} ( p' || q, H_1, H'_1)}
\\ & \\
\hline  
\multicolumn{2}{|c|}{} \\
\multicolumn{2}{|c|}{
\dedr{\cpolmsg}\sosrule{}{ (x\, \bullet \, p ; q, H, H')  \trans{x \bullet p} (  q, H, H')} ~~~\bullet \in \{?, !\}
}\\
\multicolumn{2}{|c|}{} \\
\hline 
\multicolumn{2}{|c|}{} \\
\multicolumn{2}{|c|}{
\dedr{\reconfig}\sosrule{(q, {H,  H'})  \trans{x \,\clubsuit\, p} (q', H,  H')  \quad (s, {H,  H'})  \trans{x \,\spadesuit\, p} (s', H,  H')  }{ ( q || s, H, H')  \trans{\rec{x,p}} ( q' || s' , H, H')}~~~
\begin{array}{c}
\small
     \clubsuit = ? ~~~ \spadesuit = !\\
    \textnormal{or}\\ 
     \clubsuit = ! ~~~ \spadesuit = ?
\end{array}
}
\\
\multicolumn{2}{|c|}{} \\
\hline 
\multicolumn{2}{c}{\gamma ::= (\sigma, \sigma') \mid x!q \mid x?q \mid {\rec{x,q}}} \\
\end{array} 
$
}
\end{adjustbox}
\caption{{\DNK}: Operational Semantics (relevant excerpt)}
\label{fig::sem-queue-sem} 
\end{figure*}

The operational semantics of DyNetKAT is given in Fig.~\ref{fig::sem-queue-sem}, over tuples of shape $(D,H,H')$, where $D$ is a DyNetKAT policy, $H$ is the list of packets waiting to be processed by the network, and $H'$ is the history of packets being processed according to the forwarding rules in the data plane.
Rule $\bf{(\cpolseqsucc)}$ in Fig.~\ref{fig::sem-queue-sem}, for instance, processes the current packet $\sigma$ (at the top of the waiting list) according to the NetKAT flow table encoded by $p$. The possibly modified packet is $\sigma'$, and a corresponding transition $\xrightarrow{(\sigma, \sigma')}$ can be observed in the behaviour Labelled Transition System (LTS) of the DyNetKAT model. $\sigma'$ is added to the history $H'$, and the execution of the model proceeds with the continuation $q$ and the remaining waiting packets in $H$.
Rule $\bf{(\reconfig)}$, for instance, encodes synchronous communication in DyNetKAT: a new forwarding rule or NetKAT policy $p$ is communicated via channel $x$ in a handshake between two parallel SDN components $q \Par s$ (e.g., one controller $q$ and one switch $s$). The handshake entails an execution $\xrightarrow{\rec{x,p}}$ within the DyNetKAT model. 
Rules $\bf{(\cpoloplusl)}$ and $\bf{(\intl)}$ and their symmetric counterparts define non-deterministic choice and parallel composition, respectively, in a standard fashion. Rule $\bf(\cpolX)$ simply replaces recursive variables with their definitions. Rule $\bf{(\cpolmsg)}$ encodes the axioms for asynchronous communication.
Furthermore, DyNetKAT has an ACP-like sound and complete axiomatisation for LTS bisimilarity.
A complete and thorough presentation of the DyNetKAT formal framework can be found in~\cite{CaltaisHMT22}.

\subsection{Running Example}\label{sec:run-ex}

Next, we illustrate the DyNetKAT framework by means of an example. Consider the scenario in Figure~\ref{fig:SDN-race}. A possible encoding in DyNetKAT is given in~(\ref{eq:run-ex}), as follows. 
\begin{equation}\label{eq:run-ex}
\begin{array}{rcl}
     SW & \triangleq  & (flag = {regular}) \cdot (pt = 1) \cdot (pt \leftarrow 2);  SW~\oplus \\
     & & (flag = {blocking}) \cdot (pt = 1) ; ((Help~!~1) ; SW)~\oplus \\
     & & (Up~?~1) ; SW' \\
     SW' & \triangleq & \drop; \bot \\
    C & \triangleq & (Help~?~1); ((Up~!~1);C)
\end{array}
\end{equation}

We write $(flag = regular)$ for packets {\bf not} of the blocking type. 
Whenever such a packet arrives at port 1 {$(pt = 1)$} of the switch ($SW$), it gets forwarded to port 2 {($pt \leftarrow 2$)}. Then, the switch continues recursively (denoted by $\Seq SW$).
Alternatively (denoted by $\oplus$), we write $(flag = blocking)$ to encode matching of packets of blocking type. Whenever such a packet arrives at port 1, the switch informs the controller that a new forwarding rule needs to be installed (denoted by sending the message $Help!1$). The new blocking behaviour is announced to the switch via $Up ! 1$. Upon receiving the message $Up ? 1$, the forwarding table of $SW$ is updated to $SW'$. The latter drops any incoming packet ($\drop$) and irreversibly stops from processing packets ($\bot$). The controller ($C$) repeatedly listens on channel $Help?$ for requests from the switch, and instructs the switch to install the blocking behaviour $SW'$ via $Up!1$.

\section{Vector Clocks}\label{sec:vec-clocks}
\begin{figure}[h]
    \centering
    \includegraphics[width=0.7\linewidth]{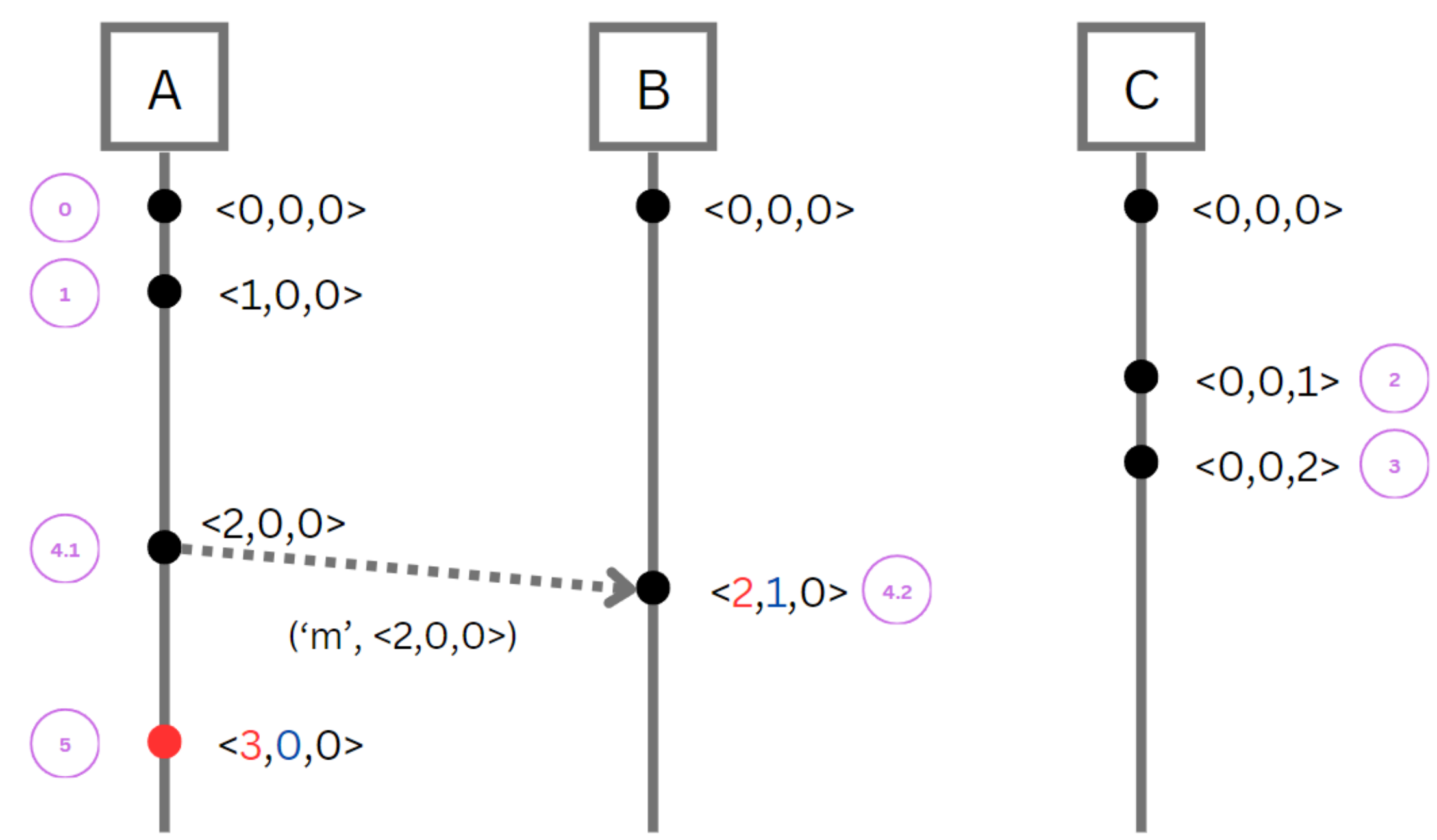}
    \caption{Vector clocks in a system with three parallel components A, B and C}
    \label{fig:vclks}
\end{figure}

SDN is a paradigm that falls under the definition of a distributed system~\cite{distributedSystems}. In this case, the components are controllers and switches, and the whole network represents a distributed system.
In distributed systems like SDNs, a data race means that switches and controllers perform actions concurrently, possibly leading to undesired behaviours. As illustrated in the introduction, for instance, there might be the case that due to concurrency, the network still forwards unsafe packets in between a new forwarding policy request, and the actual installation of the new forwarding rules. We call these \emph{data races between the control and data planes}. One possible approach to detecting data races is the use of vector clocks~\cite{DBLP:books/acm/19/Lamport19b}.
Each such clock is associated with a component in the distributed system, and it consists of a vector of size equal with the number of components in the system. Each vector entry in a clock counts actions performed by a distinct component. 
(In the context of DyNetKAT models, for instance, actions stand for packet forwarding or reconfigurations between the control and data planes.)
Each component in the system has its own copy of a vector clock as illustrated in Figure~\ref{fig:vclks}.

As shown in Figure~\ref{fig:vclks}, in step 0 all three clocks are initialized with zeros. When a component performs an individual action (i.e., no message-passing involved), it increments its own index in its local copy of the vector clock. The rest of the entries in the clock, as well as the clocks of the other components, are unchanged (see, e.g., step \textcircled{1} or \textcircled{5} of component A). Both steps \textcircled{2} and \textcircled{3} correspond to a similar scenario, but within component C.

Synchronous communication is handled as follows:
Once a component sends a message, it first increments its clock, and then sends it along with the message, creating a timestamped message (step \textcircled{4}.1). Upon its arrival, the receiver updates the rest of the clock entries in the local copy if the corresponding entries in the message timestamp are greater, and then it increments its clock entry (step \textcircled{4}.2). Note that step 4 consists of two parts capturing the synchronous sending and receiving of a message in one time frame (i.e., caputers a handshake communication).

To understand how vector clocks help detect data races we first need to know what does it mean for two vector clocks to be comparable.
Consider two vector clocks \begin{math}V_i\end{math} and \begin{math}V_j\end{math} of size $k$. The clocks are comparable if:
\begin{displaymath}
    V_i[x] \leq V_j[x] \qquad \forall x \in \{1, \ldots, k\}
\end{displaymath}
or
\begin{displaymath}
    V_i[x] \geq V_j[x] \qquad \forall x \in  \{1, \ldots, k\}
\end{displaymath}
Such comparable pair of vector clocks indicates that the associated components did not run concurrently. If there exist $x,\,y \in \{1, \ldots, k\}$ with $x \neq y$ such that:
\begin{displaymath}
    (V_i[x] \not \leq V_j[x] \quad \land \quad V_i[y] \not \geq V_j[y]) 
\end{displaymath}
then we conclude that components associated with $V_i$ and $V_j$ operate concurrently, implying a data race.
Steps \textcircled{5} and \textcircled{4}.2 for instance, witness concurrent behaviour between A and B.
Similarly for \textcircled{5} and \textcircled{2}, etc.

\section{Overview of Symbolic DyNetKAT}\label{sec:symb-dynetkat}

In this section, we briefly recall the contribution in~\cite{SDN-CHF} that introduces a symbolic operational semantics of DyNetKAT, enriched with vector clocks for detecting races between the control and data planes. The most important idea behind the symbolic DyNetKAT reduces to exploiting the so-called DyNetKAT \emph{head normal forms} that enable simulating packet processing within SDN models in a purely syntactic fashion, without actual packets being ``fed'' to the network.

The idea is as follows: each (guarded) DyNetKAT policy $d$ can be equivalently expressed (based on its complete axiomatisation in~\cite{CaltaisHMT22}) as a sum $(~\oplus~)$ of DyNetKAT policies of shape $\alpha \cdot \pi \Seq d'$ or $\rec{x,n} \Seq d'$. Here, $\alpha$ stands for a so-called \emph{complete test} $(f_1 = v_1) \cdot \ldots \cdot (f_n = v_n)$ encoding all the conditions an incoming packet has to match within a flow table, in order to be forwarded accordingly. Each packet passing a complete test as before is, in fact, a packet of shape $\sigma_{\alpha} \triangleq \{ f_1 = v_1,  \ldots, f_n = v_n\}$; so, a complete test encodes an incoming packet. A \emph{complete assignment} $\pi$ as before, is a policy $(f_1 \leftarrow v'_1) \cdot \ldots \cdot (f_n \leftarrow v'_n)$ encoding how the packet matching the complete test is processed by the data plane. Basically, a complete assignment encodes a forwarded/processed packet $\sigma_{\pi} \triangleq \{f_1 = v'_1,  \ldots, f_n = v'_n \}$.
It is, therefore, easy to understand that the symbolic semantics of DyNetKAT can be defined based on such normal forms which entail transitions of shape $\xrightarrow{(\sigma_{\alpha}, \sigma_{\pi} )}$ and $\xrightarrow{\rec{x,n}}$, respectively, without the need of actual packets.

In~\cite{SDN-CHF}, each SDN encoding a set of parallel switches ($S_i$) and controllers ($C_j$)
\begin{equation}\label{eq:res-SDN}
    S_1 \Par \ldots \Par S_n \Par C_1 \Par \ldots \Par C_m
\end{equation}
is enriched with vector clocks $\vec{c}_k$ associated with each component
 \begin{equation}\label{eq:symb-state}
    {S_1}_{\vec{c_1}} \Par \ldots \Par {C_m}_{\vec{c_m}}
 \end{equation}
entailing DyNetKAT symbolic operational rules. For instance:
\[
\dedr{\Symb{{\checkmark}}} \sosrule{p_i \in \NetKATnoDup ~~~~~  n.f.(p_i) = \Sigma_{\alpha_i \cdot \pi_i \in {\cal A}} \alpha_i \cdot \pi_i}
{ (p_i ; q_i)_{\vec{c_i}} \Par \Pi_{\begin{array}{c} 1 \leq j \leq n \\ j \not = i\end{array}} {d_j}_{\vec{c_j}}  \trans{(\sigma_{\alpha_i}, \sigma_{\pi_i})} ({q_i})_{\vec{c_i}[i]\!+\!+} \Par \Pi_{\begin{array}{c} 1 \leq j \leq n \\ j \not = i\end{array}} {d_j}_{\vec{c_j}}}
\]
is the symbolic counterpart of $\bf{(\cpolseqsucc)}$, where the vector clock $\vec{c}_i$ of the ``evolving'' component $p_i \Seq q_i$ is incremented in accordance with the semantics of the vector clocks in Section~\ref{sec:vec-clocks}, and the input packet $\sigma_{\alpha_i}$ and the processed packet $\sigma_{\pi_i}$ defining this step $\xrightarrow{(\sigma_{\alpha_i}, \sigma_{\pi_i})}$ are entailed based on the normal form of $p_i$ (note that normal forms exist for NetKAT as well~\cite{netkat}).

Here we write:
\[
{P_i}_{\vec{c_i}} \Par \Pi_{\begin{array}{c} 1 \leq j \leq k \\ j \not = i\end{array}} {P_j}_{\vec{c_j}}
\]
to denote
\[
{P_1}_{\vec{c_1}} \Par \ldots \Par {P_k}_{\vec{c_k}}
\]

The symbolic rule for handshake (i.e., the counterpart of $\bf{(\intl)}$) is defined is a similar fashion, where both vector clocks of the communicating SDN components are updated, and the transition step is marked as $\xrightarrow{\rec{x,p}}$:
\[
\dedr{\Symb{\Par}}\sosrule{
h.n.f(q_i) \triangleq x!q;d_i \oplus r_i
~~~
h.n.f(q_k) \triangleq x?q;d_k \oplus r_k
}{
\begin{array}{c}
(q_1)_{\vec{c_1}} \Par \ldots \Par (q_i)_{\vec{c_i}} \Par \ldots 
\Par (q_k)_{\vec{c_k}} \Par \ldots 
\Par (q_n)_{\vec{c_n}}\\
\trans{\rec{x,q}}\\
(q_1)_{\vec{c_1}} \Par \ldots \Par (d_i)_{\vec{c_i}[i]++} \Par \ldots 
\Par (d_k)_{(max(\vec{c_i}[i]++,\vec{c_k}))[k]++} \Par \ldots 
\Par (q_n)_{\vec{c_n}}
\end{array}
}
\]

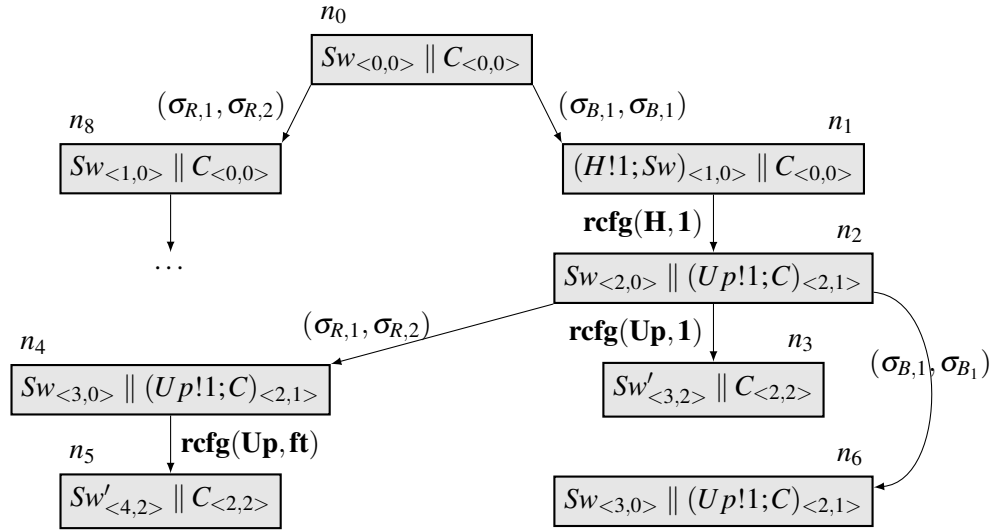
\begin{figure}
  \centering
  \begin{tikzpicture}[remember picture,
    block/.style={
      rectangle,
      draw,
      thick,
      fill=gray!20,
      align=center,
      minimum height=10pt,
    }]

   \node (n0) [label={[anchor=south west]north west:$n_0$},block] {$Sw_{<0,0>}\parallel C_{<0,0>}$};
     
   \node (n1) [below right=2em and 1em of n0, label={[anchor=south east]north east:$n_1$},block] {$(H!1;Sw)_{<1,0>}\parallel C_{<0,0>}$};
   
   \node (n2) [below=2em of n1, label={[anchor=south east]north east:$n_2$},block] {$Sw_{<2,0>}\parallel (Up!1;C)_{<2,1>}$};
   
   \node (n3) [below=2em of n2, label={[anchor=south east]north east:$n_3$},block] {$Sw'_{<3,2>}\parallel C_{<2,2>}$};

   \node (n6) [below=2em of n3, label={[anchor=south east]north east:$n_6$},block] {$Sw_{<3,0>}\parallel (Up!1;C)_{<2,1>}$};
   
   \node (n8) [below left=2em and 1em of n0, label={[anchor=south west]north west:$n_8$},block] {$Sw_{<1,0>}\parallel C_{<0,0>}$};
   
   \node (n81) [below=2em of n8]{$\cdots$};
   
   \node (n4)  at (n81 |- n3) [label={[anchor=south west]north west:$n_4$},block] {$Sw_{<3,0>}\parallel (Up!1;C)_{<2,1>}$};

   \node (n5) [below=2em of n4, label={[anchor=south west]north west:$n_5$},block] {$Sw'_{<4,2>}\parallel C_{<2,2>}$};

   \draw[-latex] (n0.south west) -- node[left, yshift=3pt]{$(\sigma_{R,1},\sigma_{R,2})$} (n8.north east);
   \draw[-latex] (n0.south east) -- node[right, yshift=3pt]{$(\sigma_{B,1},\sigma_{B,1})$} (n1.north west);
   \draw[-latex] (n1) -- node[left]{$\rec{H,1}$} (n2);
   \draw[-latex] (n8) -- (n81);
   \draw[-latex] (n2.south west) -- node[left, yshift=3pt]{$(\sigma_{R,1},\sigma_{R,2})$} (n4.north east);
   \draw[-latex] (n2) -- node[left]{$\rec{Up,1}$} (n3);
   \draw[-latex] (n2) to [bend left=85] node[midway, above] {$(\sigma_{B,1}, \sigma_{B_1})$} (n6);
   \draw[-latex] (n4) -- node[right]{$\rec{Up,ft}$} (n5);
   
  \end{tikzpicture}  
  \caption{Symbolic execution of the SDN in~(\ref{eq:run-ex}); excerpt}
  \label{fig:sym-race}
\end{figure}

Figure~\ref{fig:sym-race} illustrates the symbolic execution of the SDN in~(\ref{eq:run-ex}). 
For brevity of notation, we write: $\sigma_{B,1}$ to denote a packet $\{flag = blocking, pt = 1\}$, $\sigma_{R,1}$ to encode a packet $\{flag = regular, pt = 1\}$ and $\sigma_{R,2}$ in lieu of $\{flag = regular, pt = 2\}$.
As intuitively explained in Section~\ref{sec:intro}: if Host 1 starts sending blocking traffic to the switch on port $1$, a data race may occur. The race arises because the outcome for a new packet depends on the timing. The new packet will either be (i) forwarded according to the existing forwarding policy installed in the switch, if it arrives before the blocking rule from the controller, or (ii) it will be dropped if the blocking rule is in place first. Case (i) matches the symbolic execution $n_0 \rightarrow n_1 \rightarrow n_2 \rightarrow n_4$: instead of immediately installing the ``drop everything'' policy in $SW'$, the network first forwards a regular packet from port 1 to port 2.
The race is detected by the incomparable clocks $\langle 3, 0 \rangle$ and $\langle 2, 1 \rangle$ in $n_4$. Furthermore, the sequence associated packets $\sigma_{B,1}$ and $\sigma_{R,1}$ can be seen as the (minimal) explanation of the race.
A similar reasoning holds for the race in $n_0 \rightarrow n_1 \rightarrow n_2 \rightarrow n_6$.
Case (ii) corresponds to the symbolic execution $n_0 \rightarrow n_1 \rightarrow n_2 \rightarrow n_3$. Note that all vector clocks can be compared along this execution, so no data race is identified.

\section{Tracer}\label{sec:tracer}
Tracer~\cite{tracer} is the tool developed in this work. It exploits the symbolic semantics of DyNetKAT as described in Section~\ref{sec:symb-dynetkat}, and computes minimal sets of packets that enable races between the control and data planes of an inputted SDN encoded in DyNetKAT. In this section we provide the algorithm behind Tracer, instructions on how to install and run the tool, and one example of using Tracer.

We define a \underline{r}ace \underline{d}etection function $rd(SDN,k)$ in~(\ref{eq:race-detect}) that identifies minimal symbolic executions of SDN witnessing races up to a given depth $k$ in the execution tree of SDN (as illustrated in Section~\ref{sec:symb-dynetkat}). Furthermore, the function returns the network packets enabling these races, as explanations. The traced packets are encoded as $\alpha^i$ in~(\ref{eq:race-detect}).(d). Recall that every complete test $\alpha_i = (f_1 = v_1) \cdot \ldots \cdot (f_m = v_m)$ entails a unique packet $\{f_1 = v_1, \ldots, f_m = v_m\}$. The Tracer Algorithm~\ref{ref:alg} implements the $rd(-)$ function based on an interplay behind Python and Maude. Invoking Maude is for deriving DyNetKAT policies in head normal forms according to the DyNetKAT complete axiomatization in~\cite{CaltaisHMT22}. These head normal forms (denoted by $hnf(d_i)$ and $hnf(d_j)$ in~(\ref{eq:race-detect}).(d) and in~(\ref{eq:race-detect}).(e)) are further exploited for identifying packet forwarding steps within the analysed SDN ($\alpha^i \cdot \pi_{\alpha^i}$ in~(\ref{eq:race-detect}).(d)), or communication steps between the data and control planes ($\rec{X,p}$ based on $X_{\gamma^i}!p_{\gamma^i}$ and $X_{\gamma^j}?p_{\gamma^j}$ in~(\ref{eq:race-detect}).(e)). Observe that vector clocks are updated in accordance with the clocks of the symbolic rules: (\ref{eq:race-detect}).(d) complies to $(\Symb{{\checkmark}})$ and (\ref{eq:race-detect}).(e) complies to $(\Symb{{\Par}})$, respectively. The function $\textit{not-race}(\vec{C_1}, \ldots, \vec{C_n})$ returns {\bf true} whenever any two vector clocks $\vec{C_i}$ and $\vec{C_j}$ are incomparable, and {\bf false} otherwise.

\begin{equation}\label{eq:race-detect}
\begin{array}{r l}
(a) & rd(\Pi_{1 \leq i \leq n} d_{i_{\vec{C_i}}}, 0)  \triangleq  \emptyset\\
(b) & rd(\Pi_{1 \leq i \leq n} d_{i_{\vec{C_i}}}, k+1) \triangleq\\
(c) & \quad \textit{if not-race}(\vec{C_1}, \ldots, \vec{C_n})~\textit{then return}\\
(d) & \quad \quad \bigcup_{\substack{ 1 \leq i \leq n \\ (\alpha^i \in A^i) \land (\alpha^i \cdot \pi_{\alpha^i} \Seq d_{\alpha^i} \in hnf(d_i))}} \alpha^i :: rd(d_{\alpha^i_{\vec{C_i}[i]++}} \Par \Pi_{\substack{
     1 \leq j \leq n  \\
     j \not = i}} d_{j_{\vec{C_j}}}, k)\\
(e) & \quad \quad \bigcup_{\substack{ 1 \leq i \not = j \leq n \\
(\gamma^i \in \Gamma^i) \land (X_{\gamma^i}!p_{\gamma^i}\Seq d_{\gamma^i} \in hnf(d_i))\\
(\gamma^j \in \Gamma^j) \land (X_{\gamma^j}?p_{\gamma^j}\Seq d_{\gamma^j} \in hnf(d_j))\\
(X_{\gamma^i} == X_{\gamma^j} == X) \land (p_{\gamma^i} == p_{\gamma^j} == p)
}}
\rec{X,p} :: rd(d_{{\gamma^i}_{~\vec{C_i}[i]++}} 
\Par d_{{\gamma^j}_{~max(\vec{C_i}[i]++,\vec{C_j})[j]++}}
\Par \Pi_{\substack{
     1 \leq j \leq n  \\
     j \not = i}} d_{j_{\vec{C_j}}}, k)\\
(f) & \textit{else return } \downarrow
\end{array}
\end{equation}

We use $\downarrow$ in~(\ref{eq:race-detect}).(f) as a marker symbol indicating that $rd(-)$ identified a race witnessing trace. Every trace ending with $\downarrow$ returned by $rd(SDN,k)$ encodes a set of packets witnessing concurrent behaviour within the SDN. We use $::$ in~(\ref{eq:race-detect}) as a constructor (concatenation) for such witnesses.

From an algorithmic perspective: (\ref{eq:race-detect}).(d) is handled in lines $13-21$ of Algorithm~\ref{ref:alg}, whereas
(\ref{eq:race-detect}).(e) is handled in lines $22-31$.
Note that the aforementioned head normal forms in (\ref{eq:race-detect}) are computed using the DyNetKAT axiomatization implemented in Maude~\cite{CaltaisHMT22}: lines $8, 15, 24$ and $25$ in Algorithm~\ref{ref:alg}.
Line $8$ invokes the application of a Maude-defined ``projection'' operator $\texttt{pi\{m}\}$ that unfolds the given expression $N_i$ up to depth $\texttt{m}$.
Checking for deadlock in line $12$ of Algorithm~\ref{ref:alg} is a stopping condition based on whether all parallel components in $\texttt{curr}$ are either $\bot$ or start with communication actions that cannot be matched by any other component.
Lines $33-34$ extract the race witnessing packets, in a post-processing step.

\begin{algorithm}
\caption{Detecting races in SDN using DyNetKAT models}\label{ref:alg}
\begin{algorithmic}[1]

\STATE {\bf Input:} SDN with $k$ components as a DyNetKAT model $(N_1 \parallel N_2 \parallel \ldots \parallel N_k)$
\STATE {\bf Input:} Depth $m$ of the search
\STATE {\bf Output:} The smallest sets of network packets enabling races in the SDN 

\FOR{$i \in \{1, \ldots, k\}$}
    \STATE Initialize vector clocks $C_i = \langle 0, \ldots, 0 \rangle$ of size $k$
\ENDFOR

\FOR{$i \in \{1, \ldots, k\}$}
    \STATE Let $pmN_i$ be the result of invoking Maude \texttt{> reduce} $pi\{m\}(N_i)$
\ENDFOR

\STATE Initialize \texttt{curr} with $(pmN_1, C_1) \parallel (pmN_2, C_2) \parallel \ldots \parallel (pmN_k, C_k)$
\STATE Initialize \texttt{symb-traces} with $\emptyset$

\IF{\texttt{not-deadlock(curr)}}
    \FOR{each element $(N_i, C_i)$ at position $i$ in \texttt{curr}}
        \IF{$(\text{complete-test-assignment} ; d_i)$ is a summand of $N_i$}
            \STATE Let $rd_i$ be the result of invoking Maude \texttt{> reduce} $d_i$
            \STATE Set \texttt{curr} to $(N_1, C_1) \parallel \ldots \parallel (rd_i, C_i') \parallel \ldots \parallel (N_k, C_k)$ 
            \STATE where $C_i'$ is $C_i$ incremented at position $i$
            \STATE Append $(\textnormal{complete-test}, (C_1, \ldots, C_i', \ldots, C_k))$ to \texttt{symb-traces}
            \STATE \textbf{go to} step 12
        \ENDIF
    \ENDFOR

    \FOR{all pairs of elements $(N_i, C_i)$ and $(N_j, C_j)$ at positions $i$ and $j$ in \texttt{curr}}
        \IF{$(X!p ; d_i)$ is a summand of $C_i$ and $(X?p ; d_j)$ is a summand of $C_j$}
            \STATE Let $rd_i$ be the result of invoking Maude \texttt{> reduce} $d_i$
            \STATE Let $rd_j$ be the result of invoking Maude \texttt{> reduce} $d_j$
            \STATE Set \texttt{curr} to $(N_1, C_1) \parallel \ldots \parallel (rd_i, C_i') \parallel \ldots \parallel (rd_j, C_j') \parallel \ldots \parallel (N_k, C_k)$ 
            \STATE where $C_i'$ is $C_i$ incremented at position $i$, and $C_j'$ is $\max(C_i', C_j)$ incremented at position $j$
            \STATE Append $(\texttt{rcfg}(X, p), (C_1, \ldots, C_i', \ldots, C_j', \ldots, C_k))$ to \texttt{symb-traces}
            \STATE \textbf{go to} step 12
        \ENDIF
    \ENDFOR
\ENDIF

\STATE Let \texttt{races} be a set of sets of packets, initialized with $\emptyset$

\FOR{all \texttt{symb-trace} in \texttt{symb-traces}}
    \IF{a prefix \texttt{s-tr'} of \texttt{symb-trace} ends with incomparable vector clocks $(C_1, \ldots, C_k)$}
        \STATE Extract all packets \texttt{pkt} based on every \texttt{max-test} in \texttt{s-tr'}
        \STATE Add the set of packets \texttt{pkt} to \texttt{races}
    \ENDIF
\ENDFOR

\STATE \Return \texttt{races}

\end{algorithmic}
\end{algorithm}

\bigskip
\noindent{\bf Tracer at Work.}
Tracer is publicly available at~\cite{tracer}. A complete installation guide can be found in README.md. Requirements for running Tracer include a Linux operating system, specifically Ubuntu 20.04\footnote{Other Linux distributions might work, however, the development and testing were done on the specified version of Ubuntu.} with Python (version $>3.10.12$). The tool also uses Maude 3.1, that is included in the installation of Tracer. To use it, run the command in the following form:
\begin{verbatim}
> python tracer\_runner.py <path_to_maude> <path_to_model_in_maude>
\end{verbatim}
The command has several optional parameters as given in Table~\ref{tab:runneropts}. Note that the parameters with values should be inputted without the space between the parameter and the value. For example, write \verb|-grace| to produce only  graphs and traces witnessing data races in the provided model.

\begin{table}[H]
\centering \caption{Tracer command line}
\resizebox{\columnwidth}{!}{
    \begin{tabular}{|c|c|l|} \hline
        \textbf{Parameter}                  & \textbf{Value} & \textbf{Explanation}\\ \hline 
        -c             & - & output text with color \\ \hline 
        -t    & - & show tracing steps \\ \hline
        -u    & int & unfold depth \\ \hline
        -g    & `race' or `full' & types of trees and traces to generate (race witnesses only, or full trees/traces) \\ \hline
        -f    & string & set a name for text output file (copy of console output)\\ \hline
    \end{tabular}
}
\label{tab:runneropts}
\end{table}

\bigskip
\noindent{\bf SDN Encoding in Tracer: Example.}
The DyNetKAT encoding in~(\ref{eq:run-ex}) is provided as input for Tracer in a Maude-compatible format as shown in Listing~\ref{lst:maude}. The DyNetKAT recursive variables $SW, SW'$ and $C$ in~(\ref{eq:run-ex}) are declared as the constants $\texttt{SW}, \texttt{SWP}$ and $\texttt{C}$ of $\texttt{Recursive}$ type in Listing~\ref{lst:maude}.
The operator $\texttt{getRecPol(...)}$ is a syntactic wrapper around these recursive operators.
The actual definitions of the switch and controller follow closely the syntax in~(\ref{eq:run-ex}).
The communication channels $Help$ and $Up$ translate to the constants $\texttt{Help}$ and $\texttt{Up}$ of type $\texttt{Channel}$ in Maude. The DyNetKAT non-deterministic choice $\oplus$ translates to $\texttt{o+}$ in Maude. The DyNetKAT constants $\drop$ and $\bot$ are mapped to $\texttt{zero}$ and \texttt{bot}. The entire SDN consisting of the switch $SW$ and controller $C$ as in Figure~\ref{fig:SDN-race} is defined by the constant $\texttt{Init}$ of type $\texttt{DNA}$ in Maude. Note that the NetKAT expressions encoding the forwarding policies are provided as strings in Maude; e.g., $\texttt{"(flag = regular).(pt = 1).(pt <- 2)"}$.
The model in Listing~\ref{lst:maude} along with the depth $k$ of the analysis are provided as input to Tracer.

\lstset{
  basicstyle=\small\ttfamily, 
  showstringspaces=false,
  tabsize=2,
  breaklines=true,
  captionpos=b,
  frame=single,
  morekeywords={fmod, protecting, ops, eq, endfm} 
}

\begin{lstlisting}[language=,caption={Maude encoding of the SDN in~(\ref{eq:run-ex})},label={lst:maude}]
fmod MODEL is
    [...]

    ops Init : -> DNA .
    ops SW, SWP, C : -> Recursive .
    ops Help, Up : -> Channel .

    eq getRecPol(SW) =
         "(flag = regular) . (pt = 1) . (pt <- 2)" ; SW o+
         "(flag = blocking) . (pt = 1) . 1" ;( (Help ! "one") ; SW ) o+
         (Up ? "one") ; SWP .
    eq SWP = zero ; bot .
    eq getRecPol(C) = (Help ? "one") ; ( (Up ! "one") ; C ) .

    eq Init = C || SW .
endfm
\end{lstlisting}

\begin{figure}[H]
    \centering
\begin{tikzpicture}
    \node (n0) at (0,0) [circle,draw,fill=green!20,minimum size=1cm,inner sep=0] 
    {${\begin{array}{c}
       0  \\
       C_{\langle 0, 0\rangle} \Par
       SW_{\langle 0, 0\rangle}
    \end{array}}$};
    \node (n1) at (4,0) [circle,draw,fill=blue!20,minimum size=1cm,inner sep=0] {${\begin{array}{c}
       1  \\
       C_{\langle 0, 0\rangle} \Par
       SW_{\langle 0, 1\rangle}
    \end{array}}$};
    \node (n2) at (8,0) [circle,draw,fill=yellow!20,minimum size=1cm,inner sep=0]
    {${\begin{array}{c}
       3  \\
       C_{\langle 1, 2\rangle} \Par
       SW_{\langle 0, 2\rangle}
    \end{array}}$};
    \node (n3) at (12,3) [circle,draw,fill=purple!20,minimum size=1cm,inner sep=0]
    {${\begin{array}{c}
       5  \\
       C_{\langle 1, 2\rangle} \Par
       SW_{\langle 0, 3\rangle}
    \end{array}}$};
    \node (n4) at (12,-3) [circle,draw,fill=purple!20,minimum size=1cm,inner sep=0]
    {${\begin{array}{c}
       6  \\
       C_{\langle 1, 2\rangle} \Par
       SW_{\langle 0, 3\rangle}
    \end{array}}$};

    \draw[->] (n0) -- (n1) node[midway,above,rotate=90] {($\{fl=B, pt=1\},{\{fl=B, pt=1\}}$)};
    \draw[->] (n1) -- (n2) node[midway,above,rotate=90] {$\rec{H,1}$};
    \draw[->] (n2) -- (n3) node[midway,above,rotate=-45] {(\{fl=B, pt=1\},{\{fl=B, pt=1\}})};
    \draw[->] (n2) -- (n4) node[midway,above,rotate=45] {(\{fl=R, pt=1\},{\{fl=R, pt=2\}})};
\end{tikzpicture}

    \caption{Races in $SW ~|| ~C$ up to depth $3$}
    \label{fig:tree3}
\end{figure}
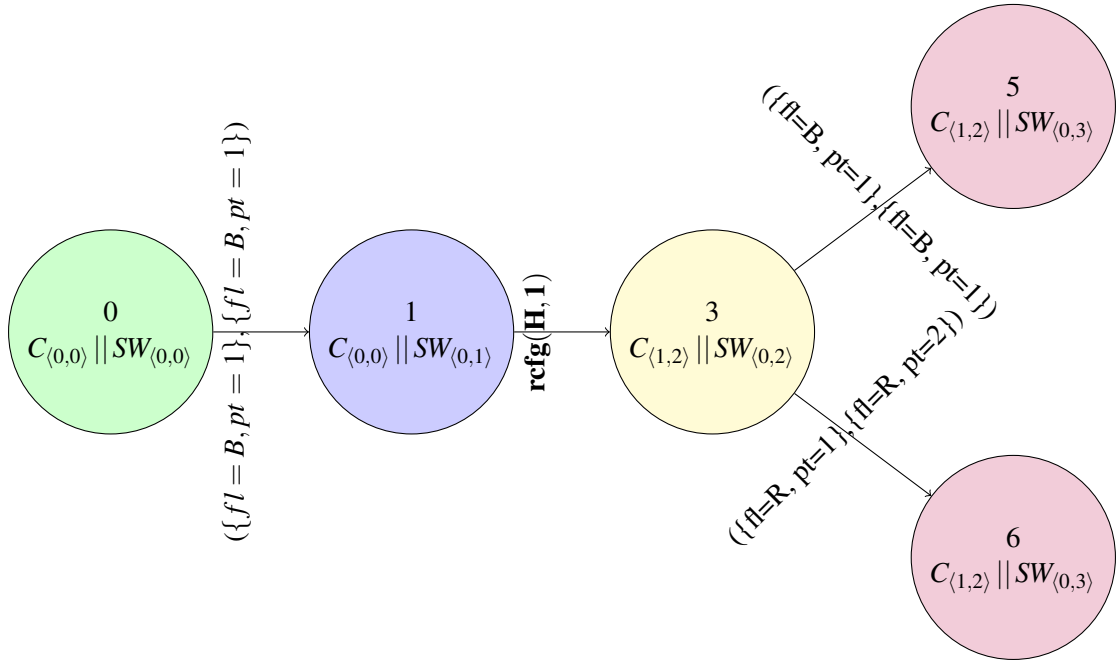

\noindent{\bf Tracer Output: Example.}
Figure~\ref{fig:tree3} showcases the output races as identified by Tracer, in a graphical format.
(We use $fl, B$ and $R$ as shorthand for $flag, blocking$ and $regular$, respectively.)
The sequence of nodes $0 \rightarrow 1 \rightarrow 3 \rightarrow 5$ corresponds to the symbolic execution $n_0 \rightarrow n_1 \rightarrow n_2 \rightarrow n_4$ in Figure~\ref{fig:sym-race}, encoding a race.
The sequence of nodes $0 \rightarrow 1 \rightarrow 3 \rightarrow 6$ in Figure~\ref{fig:tree3} corresponds to the symbolic execution $n_0 \rightarrow n_1 \rightarrow n_2 \rightarrow n_6$ Figure~\ref{fig:sym-race}, encoding a race as well. Furthermore, the labels along these executions are minimal explanations of how the races can be enabled. Note how the corresponding clocks in Figure~\ref{fig:tree3} match their counterparts in Figure~\ref{fig:sym-race}. 

Tracer outputs the witnesses of data races in a textual format as well, as illustrated in Figure~\ref{fig:tracesu3}.
These traces can be in short form, encoding the input packets  and/or reconfiguration steps within symbolic executions without vector clocks. Traces in long form show the action performer (switch \texttt{SW}, controller \texttt{C} or a handshake between the switch and the controller \texttt{SW -> C}), the vector clocks (\texttt{[0, 0]}, \ldots), and the corresponding node ID in the graph as well. 

\lstset{ 
  basicstyle=\footnotesize\ttfamily, 
  frame=single,               
  captionpos=b,               
  numbers=left,               
  numberstyle=\tiny,          
}

\begin{figure*}
    \centering
    \begin{minipage}{\textwidth}
    \begin{lstlisting}
RACE SHORT TRACES
Trace 0:
"(flag = blocking) . (pt = 1) "; rcfg('Help', '"one"');
"(flag = blocking) . (pt = 1)"

Trace 1:
"(flag = blocking) . (pt = 1) . 1"; rcfg('Help', '"one"');
"(flag = regular) . (pt = 1)"



RACE LONG TRACES
Trace 0:
{C[0, 0] || SW[0, 0]} nid:0;
[SW] "(flag = blocking) . (pt = 1) " {C[0, 0] || SW[0, 1]} nid:1;
[SW -> C] rcfg('Help', '"one"') {C[1, 2] || SW[0, 2]} nid:3;
[SW] "(flag = blocking) . (pt = 1)" {C[1, 2] || SW[0, 3]} nid:5;


Trace 1:
{C[0, 0] || SW[0, 0]} nid:0;
[SW] "(flag = blocking) . (pt = 1) " {C[0, 0] || SW[0, 1]} nid:1;
[SW -> C] rcfg('Help', '"one"') {C[1, 2] || SW[0, 2]} nid:3;
[SW] "(flag = regular) . (pt = 1) " {C[1, 2] || SW[0, 3]} nid:6;
    \end{lstlisting}
    \caption{Race traces of $SW ~|| ~C$ with unfold 3}
    \label{fig:tracesu3}
    \end{minipage}
\end{figure*}

\section{Conclusions}\label{sec:conc}

In this paper, we introduced Tracer~\cite{tracer}, a tool for detecting and explaining data races in SDNs as defined in~\cite{SDN-CHF}. These systems exhibit concurrent behavior due to the interaction between data plane processing, and dynamic reconfigurations between the data and control planes.
Tracer focuses on pin-pointing data races in SDN models encoded within the DyNetKAT~\cite{CaltaisHMT22} framework.
In addition, Tracer provides explanations of how these data races can be enabled by identifying sequences of packets which, whenever fed to the SDN under analysis, lead to concurrency between the data and control planes.
The tool is built on top of the DyNetKAT axiomatisation implemented in Maude. In the future, we plan to analyze Tracer's performance on benchmarks with larger SDN models. Additionally, we aim to implement a parallelized version of Tracer to improve its efficiency.

\medskip
\noindent
{\bf Acknowledgements.} {This work was supported by the project ZORRO, no. KICH1.ST02.21.003 of the research programme Key Enabling Technologies (KIC) which is (partly) financed by the Dutch Research Council (NWO).}

\clearpage

\end{document}